\def\rfr#1{eq. (\ref{#1})}

\def\dert#1#2{\frac{{{d}}{#1}}{{{d}}{#2}}}              % derivate parziali e totali prima e seconda

\def\virg#1{``#1''}

\def\bb#1#2#3{\bibitem[\protect\citeauthoryear{#1}{#2}]{#3}}
\def\eqi{\begin{equation}}
\def\eqf{\end{equation}}
\def\eqia{\begin{eqnarray}}
\def\eqfa{\end{eqnarray}}
\def\rp#1#2{{#1\over#2}} \def\lb#1{\label{#1}}

\def\bds#1{\boldsymbol{#1}}

\documentclass[usenatbib]{mn2e}
\usepackage{amsmath}
\usepackage{amscd}
\usepackage{amssymb}
\usepackage{latexsym}
\usepackage{graphicx}
\usepackage{epsfig}

\title[Constraints on planet X/Nemesis from Solar System's inner dynamics]{Constraints on planet X/Nemesis from Solar System's inner dynamics}

\author[L. Iorio]{
L. Iorio$^{1}$\thanks{E-mail:
lorenzo.iorio@libero.it}\\
$^{1}$INFN-Sezione di Pisa, Viale Unit$\grave{\rm a}$ di Italia 68, 70125, Bari (BA), Italy
}

\begin{document}

%\date{Accepted 1988 December 15. Received 1988 December 14; in original form 1988 October 11}

%\pagerange{\pageref{firstpage}--\pageref{lastpage}} \pubyear{2002}

\maketitle

\label{firstpage}

\begin{abstract}
 We use the corrections $\Delta\dot\varpi$ to the standard Newtonian/Einsteinian  perihelion precessions of the inner planets of the solar system, recently estimated by E.V. Pitjeva by fitting a huge planetary data set with the dynamical models of the EPM ephemerides, to put constraints on the position of a putative, yet undiscovered large body X of mass $M_{\rm X}$, not modelled in the EPM software. The direct action of X on the inner planets can be approximated by a Hooke-type radial acceleration plus a term of comparable magnitude having a fixed direction in space pointing towards X. The perihelion precessions induced by them can be analytically worked out only for some particular positions of X in the sky; in general, numerical calculations are used. We show that the indirect effects of X on the inner planets through its action on the outer ones can be neglected, given the present-day level of accuracy in knowing $\Delta\dot\varpi$. As a result, we find that Mars yields the tightest constraints, with the tidal parameter ${\mathcal{K}}_{\rm X}= GM_{\rm X}/r_{\rm X}^3\leq 3\times 10^{-24}$ s$^{-2}$. To constrain $r_{\rm X}$ we  consider the case of a rock-ice planet with the mass of Mars and the Earth, a giant planet with the mass of Jupiter, a brown dwarf with $M_{\rm X}=80m_{\rm Jupiter}$, a red dwarf with $M=0.5M_{\odot}$ and a Sun-mass body. For each of them we plot $r^{\rm min}_{\rm X}$ as a function of the heliocentric latitude $\beta$ and longitude $\lambda$. We also determine the forbidden  spatial region for X by plotting its boundary surface in the three-dimensional space: it shows significant departures from spherical symmetry. A Mars-sized body can be found at no less than $70-85$ AU: such bounds are  $147-175$ AU, $1,006-1,200$ AU, $4,334-5,170$ AU, $8,113-9,524$ AU, $10,222-12,000$ AU for a body with a mass equal to that of the Earth, Jupiter, a brown dwarf, red dwarf and the Sun, respectively.
 \end{abstract}

%\keywords{Solar system objects; Low luminosity stars, subdwarfs, and brown dwarfs; Kuiper belt, trans-Neptunian objects; Oort cloud; Celestial % mechanics }
%PACS\ 96.30.-t; 97.20.Vs; 96.30.Xa; 96.50.Hp; 95.10.Ce


\section{Introduction}
Does the solar system contain an undiscovered massive planet  or a distant stellar companion of the Sun?

The history of an hypothetical Planet X dates back to the early suggestions by the astronomer P. \citet{Lowell} who thought that some glitches in the orbit of Uranus  might be caused by what he dubbed Planet X. In 1930, the search that Lowell initiated led to the discovery of Pluto \citep{Tom}.
 The various constraints on the mass and position
of a putative Planet X, as candidate to accommodate the alleged orbital anomalies of Uranus \citep{coglio}, were summarized by \citet{Ho}. Later, the claimed residuals in
the orbit of Uranus were explained by \citet{Sta} in terms of small systematic errors as an underestimate  of the mass of Neptune by $0.5\%$.  However, the appeal of a yet undiscovered solar system's body of planetary size never faded. Indeed, according to \citet{quindi}, a ninth planet as large as the Earth may exist beyond Pluto, at about 100-170 AU, to explain  the architecture of the Edgeworth-Kuiper Belt. Previously, \citet{Bru} proposed the existence of a Mars-size body at 60 AU to explain certain features in the  distribution of the Trans-Neptunian Objects (TNOs) like the so-called Kuiper Cliff where low-eccentricity and low-inclination) Kuiper
Belt objects (KBOs) with semimajor axes greater than 50 AU rapidly
falls to zero, although several problems in explaining other features of the Kuiper Belt with such a hypothesis were pointed out later \citep{pro}.
According to \citet{Mate}, a perturber body of mass $m\approx 1.5m_{\rm Jup}$ at 25 kAU  would be able to explain the anomalous distribution of orbital elements of approximately $25\%$ of the 82 new class I Oort cloud comets. A similar hypothesis was put forth by \citet{Mur99}; for a more skeptical view, see \citet{Hor}.
\citet{Gomes} suggested that distant detached objects among the TNOs (perihelion distance $q>40$ AU and semimajor axis, $a>50$ AU)  may have been generated by a hypothetical Neptune-mass companion having semiminor axis $b\leq 2,000$ AU or a Jupiter-mass companion with $b\leq 5,000$ AU on significantly inclined orbits.
However, generally speaking, we stress that we still have  bad statistics about the TNOs and the Edgeworth-Kuiper Belt  objects \citep{Mor07,Schw}.

Concerning the existence of a putative stellar companion of the Sun, it was argued with different approaches. As an explanation of the peculiar properties of certain pulsars with anomalously small period derivatives, it was suggested by \citet{Har} that the barycenter of the solar system is accelerated, possibly because the Sun is a member of a binary system and has a hitherto undetected companion star.
Latest studies by \citet{Zak} constrain such a putative acceleration at a $A_{\rm bary}\lesssim 1.2\times 10^{-9}$ m s$^{-2}$ level ($4\times 10^{-18}$ s$^{-1}$ in units of $A_{\rm bary}/c$, where $c$ is the speed of light in vacuum).  \citet{Whi} and \citet{Da} suggested that the statistical periodicity of about 26 Myr in extinction rates on the Earth over the last 250 Myr reported by \citet{Rau} can be explained  by  a  yet undetected companion star (called Nemesis) of the Sun in a highly elliptical orbit that periodically would disturb comets in the Oort cloud, causing a large increase in the number of comets visiting the inner solar system with a consequential increase in impact events on Earth. In a recent work, \citet{Muller} used the hypothesis of Nemesis to explain the measurements of the ages of 155 lunar spherules from the Apollo 14 site.
The exact nature of Nemesis is uncertain; it could be a red dwarf \citep{Muller} ($ 0.075 \leq m \leq 0.5$ M$_{\odot}$), or a brown dwarf \citep{Whi} ($m \approx 75-80m_{\rm Jup}$).  For some resonant mechanisms between Nemesis and the Sun triggering  Oort cloud's comet showers  at every perihelion passage, see  \citet{Van93}. Recently, \citet{Fo} and \citet{Sil} put forth the hypothesis that Nemesis could be made up of the so called mirror matter, whose existence is predicted if parity is an unbroken symmetry of nature.

In this paper we constrain  the distance of  a very distant body for different values of its mass in a dynamical, model-independent way by looking at
the gravitational effects induced by it on the motions of the inner planets orbiting in the $0.4-1.5$ AU range. The same approach was followed by \citet{kriplo1} and \citet{kriplo2}  to put constraints on density of diffuse dark matter in the solar system. In Section \ref{calculo} we calculate the acceleration imparted by a distant body X on an inner planet P and the resulting perihelion precession averaged over one orbital revolution of P. In Section \ref{megafig} we discuss the position-dependent constraints on the minimum distance at which X can exist for several values of its mass, and depict the forbidden regions for it in the three-dimensional space. In Section \ref{fine} we compare our results to other constraints existing in literature and summarize our findings.
\section{The perihelion precessions induced by a distant massive body}\lb{calculo}
The gravitational acceleration imparted by a body of mass $M_{\rm X}$ on a planet P is, with respect to some inertial frame,
\eqi \bds A_{\rm PX} = \rp{GM_{\rm X}}{\left|\bds r_{\rm X}-\bds r_{\rm P}\right|^3}\left(\bds r_{\rm X}-\bds r_{\rm P}\right).\lb{accia}\eqf
If, as in our case, it is supposed $r_{\rm X}\gg r_{\rm P}$, by neglecting terms of order $\mathcal{O}(r_{\rm P}^2/r_{\rm X}^2)$, it is possible to use the approximated formula
\eqi \rp{1}{\left|\bds r_{\rm X}-\bds r_{\rm P}\right|^3}\approx\rp{1}{r_{\rm X}^3}\left[1+\rp{3\left(\bds r_{\rm P}\bds\cdot\bds{\hat{n}}_{\rm X}\right)}{r_{\rm X}}\right]+\mathcal{O}\left(\rp{r_{\rm P}}{r_{\rm X}}\right)^2,\lb{appr}\eqf where \eqi\bds{\hat{n}}_{\rm X}\equiv \rp{\bds r_{\rm X}}{r_{\rm X}}\eqf is the unit vector of X which can be assumed constant over one orbital revolution of P.  By inserting \rfr{appr} in \rfr{accia} and neglecting the resulting term of order $\mathcal{O}(r_{\rm P}^2/r_{\rm X}^2)$ one has
\eqi\bds A_{\rm PX} \approx -\rp{GM_{\rm X}}{r^3_{\rm X}}\bds r_{\rm P} + \rp{3GM_{\rm X}\left(\bds r_{\rm P}\bds\cdot\bds{\hat{n}}_{\rm X}\right)}{r^3_{\rm X}}\bds{\hat{n}}_{\rm X}+\rp{GM_{\rm X}}{r^2_{\rm X}}\bds{\hat{n}}_{\rm X}.\lb{final}\eqf
The acceleration of \rfr{final} consists of three terms: a non-constant radial Hooke-type term, a non-constant term directed along the fixed direction of  $\bds{\hat{n}}_{\rm X}$   and a constant (over the typical timescale of P) term having the direction of $\bds{\hat{n}}_{\rm X}$  as well.  An equation identical to \rfr{final} can be written also for the Sun by replacing everywhere $\bds r_{\rm P}$ with $\bds r_{\rm S}$; thus, since we are interested in the motion of the planet P relative to the Sun, the constant term cancels out, and by posing $\bds r\equiv \bds r_{\rm P}-\bds r_{\rm S}$ one can writes down the heliocentric perturbing acceleration felt by the planet P  due to X
\eqi\bds A_{\rm X}=\bds A_{\rm Hooke}+\bds A_{n_{\rm X}}\equiv -\rp{GM_{\rm X}}{r^3_{\rm X}}\bds r + \rp{3GM_{\rm X}\left(\bds r\bds\cdot\bds{\hat{n}}_{\rm X}\right)}{r^3_{\rm X}}\bds{\hat{n}}_{\rm X}.\lb{kazza}\eqf
In fact, the planetary observation-based quantities we will use in the following have been determined in the frame of reference of the presently known solar system's baricenter (SSB). By the way, \rfr{kazza}, where $\bds r$ is to be intended as the position vector of P with respect to the known SSB, is valid also in this case. Indeed, by repeating the same reasonings as before, we are led to a three-terms equation like \rfr{final} in which the vectors entering it refer to the SSB frame;  now, if X existed, the SSB frame would be uniformly accelerated  by \eqi \bds A_{\rm SSB} = \rp{GM_{\rm X}}{r^2_{\rm X}}\bds{\hat{n}}_{\rm X},\eqf so that P, referred to such a non-inertial SSB frame, would also be acted upon by an inertial acceleration  \eqi \bds A_{\rm in} = -\rp{GM_{\rm X}}{r^2_{\rm X}}\bds{\hat{n}}_{\rm X},\eqf in addition to the gravitational one of \rfr{final}, which just cancels out the third constant term in \rfr{final} leaving us  with  \rfr{kazza}. Note that, in general, it is not legitimate to neglect $\bds A_{n_{\rm X}}$ with respect to $\bds A_{\rm Hooke}$ since
\eqi\rp{A_{n_{\rm X}}}{A_{\rm Hooke}}=3\cos\psi,\eqf where $\psi$ is the angle between $\bds r$ and $\bds n_{\rm X}$ which, in general, may vary within $0\leq \psi\leq 2\pi$  during an orbital revolution of P. Moreover,
if the acceleration of a distant body could only be expressed by a Hooke-type term, this would lead to the absurd conclusion that no bodies at all exist outside the solar system because the existence of an anomalous Hooke-like acceleration has been ruled out by taking the ratio of the perihelia of different pairs planets   \citep{Ior}.

 The acceleration of \rfr{kazza}
is in agreement with the potential
\eqi U_{\rm X}=\rp{GM_{\rm X}}{2r_{\rm X}^3}\left[r^2 - 3\left(\bds r\bds\cdot\bds{\hat{n}}_{\rm X}\right)^2 \right]\eqf
 proposed by \citet{Ho}.

The action of \rfr{kazza} on the orbit of a known planet of the  solar system can be treated perturbatively with
the Gauss equations \citep{Ber} of the variations of the Keplerian orbital elements
{\tiny{\begin{eqnarray}\lb{Gauss}
\dert a t & = & \rp{2}{n\sqrt{1-e^2}} \left[e A_r\sin f +A_{t}\left(\rp{p}{r}\right)\right],\lb{gaus_a}\\  \nonumber \\
\dert e t  & = & \rp{\sqrt{1-e^2}}{na}\left\{A_r\sin f + A_{t}\left[\cos f + \rp{1}{e}\left(1 - \rp{r}{a}\right)\right]\right\},\lb{gaus_e}\\ \nonumber \\
\dert I t & = & \rp{1}{na\sqrt{1-e^2}}A_n\left(\rp{r}{a}\right)\cos u,\\  \nonumber \\
\dert\Omega t & = & \rp{1}{na\sin I\sqrt{1-e^2}}A_n\left(\rp{r}{a}\right)\sin u,\lb{gaus_O}\\   \nonumber \\
\dert\omega t & = &\rp{\sqrt{1-e^2}}{nae}\left[-A_r\cos f + A_{t}\left(1+\rp{r}{p}\right)\sin f\right]-\cos I\dert\Omega t,\lb{gaus_o}\\  \nonumber \\
\dert {\mathcal{M}} t & = & n - \rp{2}{na} A_r\left(\rp{r}{a}\right) -\sqrt{1-e^2}\left(\dert\omega t + \cos I \dert\Omega t\right),\lb{gaus_M}
\end{eqnarray}
}}

where $a$, $e$, $I$, $\Omega$, $\omega$ and ${\mathcal{M}}$ are the semi-major axis, the eccentricity, the inclination, the longitude of the ascending node, the argument of pericenter and the mean anomaly of the orbit of the test particle, respectively,  $u=\omega+f$ is the argument of latitude ($f$ is the true anomaly), $p=a(1-e^2)$ is the semi-latus rectum and $n=\sqrt{GM/a^3}$ is the un-perturbed Keplerian mean motion; $A_r,A_t,A_n$ are the radial, transverse and normal components of a generic perturbing acceleration $\bds A$.
By evaluating them onto the un-perturbed Keplerian ellipse
\eqi r = \rp{a(1-e^2)}{1+e\cos f}\eqf    and averaging\footnote{We used $\dot u=\dot f$ because over one orbital revolution the pericentre $\omega$ can be assumed constant.} them over one orbital period $P_{\rm b}$ of the planet  by means of
\eqi \rp{dt}{P_{\rm b}} = \rp{(1-e^2)^{3/2}}{2\pi(1+e\cos f)^2}df,\eqf
it is possible to obtain the long-period effects induced by \rfr{kazza}.

In order to make contact with the latest observational determinations, we are interested in the averaged rate of the longitude of the pericenter $\varpi\equiv \omega+\Omega$;
its variational equation is {\tiny{\eqi\dert\varpi t = \rp{\sqrt{1-e^2}}{nae}\left[-A_r\cos f + A_{t}\left(1+\rp{r}{p}\right)\sin f\right]+2\sin^2\left(\rp{I}{2}\right)\dert\Omega t.\lb{varpidot}\eqf}  }
Indeed, the astronomer  \citet{Pit05} has recently estimated, in a least-square sense, corrections $\Delta\dot\varpi$ to the standard Newtonian/Einsteinian averaged precessions of the perihelia of the inner planets of the solar system, shown in Table \ref{chebolas}, by fitting almost one century of planetary observations of several kinds with the dynamical force models of the EPM ephemerides; since they do not include the force imparted by  a distant companion of the Sun, such corrections are, in principle, suitable to constrain the unmodelled action of such a putative body accounting for its direct action on the inner planets themselves and, in principle, the indirect one on them through the outer planets\footnote{Only their standard $N-$body mutual interactions have been, indeed, modelled in the EPM ephemerides.}; we will discuss such an issue later.

Concerning the calculation of the perihelion precession induced by \rfr{kazza}, as already noted,  $\bds r_{\rm X}$ can certainly be considered as constant over the orbital periods $P_{\rm b}\lesssim 1.5$ yr of the inner planets. Such an approximation  simplifies our calculations.
 In the case of the Hook-type term of \rfr{kazza}
 \eqi \bds A_{\rm Hooke}=-\mathcal{K}_{\rm X}\bds r,\ \mathcal{K}_{\rm X}=\rp{GM_{\rm X}}{r^3_{\rm X}}\lb{hooke},\eqf the task of working out its secular orbital effects has been already performed several times in literature; see, e.g., \citet{Ior} and references therein. The result for the longitude of perihelion is
\eqi\left\langle\dot\varpi\right\rangle_{\rm Hooke}=-\rp{3}{2}\mathcal{K}_{\rm X}\rp{\sqrt{1-e^2}}{n}=-\rp{3}{2}\rp{GM_{\rm X}}{r_{\rm X}^3}\sqrt{\rp{(1-e^2)a^3}{GM_{\odot}}}.\lb{precis}\eqf
The treatment of the second term $\bds A_{n_{\rm X}}$ of \rfr{kazza} is much more complex because it is not simply radial. Indeed, its direction is given by $\bds{\hat{n}}_{\rm X}$ which is fixed in the inertial space $\{x,y,z\}$ during the orbital motion of P. Its components are the three direction cosines, so that the cartesian components of  \eqi\bds A_{n_{\rm X}}=A_x\bds i+A_y\bds j+A_z\bds k\eqf  are
\begin{eqnarray}
  A_x &=& 3{\mathcal{K}_{\rm X}}(xn_x+yn_y+zn_z)n_x, \lb{opis}\\ \nonumber\\
  A_y &=& 3{\mathcal{K}_{\rm X}}(xn_x+yn_y+zn_z)n_y, \\ \nonumber\\
  A_z &=& 3{\mathcal{K}_{\rm X}}(xn_x+yn_y+zn_z)n_z.\lb{vobis}
\end{eqnarray}
 In terms of the standard heliocentric angular coordinates\footnote{Recall that $\lambda$ is the usual longitude $\phi$ of the standard spherical coordinates, while $\beta=90^{\circ}-\theta$, where $\theta$ is the usual co-latitude of the standard spherical coordinates.} $\lambda,\beta$, we can write the components of $\bds{\hat{n}}_{\rm X}$ as
\begin{eqnarray}
  n_x &=& \sqrt{1-\sin^2\beta_{\rm X}}\cos\lambda_{\rm X}, \\ \nonumber\\
  n_y &=& \sqrt{1-\sin^2\beta_{\rm X}}\sin\lambda_{\rm X},\\ \nonumber\\
  n_z &=& \sin\beta_{\rm X},
\end{eqnarray}
with $n_x^2+n_y^2+n_z^2=1$.
In order to have the radial, transverse and normal components of $\bds A_{n_{\rm X}}$ to be inserted into the right-hand-sides of the Gauss equations
\begin{eqnarray}
  A_r &=& \bds A_{n_{\rm X}}\bds\cdot\bds{\hat{r}}, \\ \nonumber\\
  A_t &=& \bds A_{n_{\rm X}}\bds\cdot\bds{\hat{t}}, \\ \nonumber\\
  A_n &=& \bds A_{n_{\rm X}}\bds\cdot\bds{\hat{n}},
\end{eqnarray}
 we need the cartesian components of the co-moving unit vectors $\bds{\hat{r}},\bds{\hat{t}},\bds{\hat{n}}$: they are \citep{Mont}
 \eqi \bds{\hat{r}} =\left(
       \begin{array}{c}
          \cos\Omega\cos u\ -\cos I\sin\Omega\sin u\\
          \sin\Omega\cos u + \cos I\cos\Omega\sin u\\
         \sin I\sin u \\
       \end{array}
     \right)
\eqf
 \eqi \bds{\hat{t}} =\left(
       \begin{array}{c}
         -\sin u\cos\Omega-\cos I\sin\Omega\cos u \\
         -\sin\Omega\sin u+\cos I\cos\Omega\cos u \\
         \sin I\cos u \\
       \end{array}
     \right)
\eqf
 \eqi \bds{\hat{n}} =\left(
       \begin{array}{c}
          \sin I\sin\Omega \\
         -\sin I\cos\Omega \\
         \cos I\\
       \end{array}
     \right)
\eqf

 %
 %{\footnotesize{\begin{eqnarray}
%\bds{\hat{r}} &=& (\cos\Omega\cos u\ -\cos I\sin\Omega\sin u)\ \bds i + (\sin\Omega\cos u + \cos I\cos\Omega\sin u)\ \bds j + \sin I\sin u\ \bds k, \\  %\nonumber \\
  %
%  \bds{\hat{t}} &=& (-\sin u\cos\Omega-\cos I\sin\Omega\cos u)\ \bds i+(-\sin\Omega\sin u+\cos I\cos\Omega\cos u)\ \bds j + \sin I\cos u\ \bds k, \\  %\nonumber  \\
  %
 % \bds{\hat{n}} &=& \sin I\sin\Omega\ \bds i-\sin I\cos\Omega\ \bds j + \cos I\ \bds k.
%\end{eqnarray}  } }
%
In our specific case, $\Omega$ is the longitude of the ascending node which yields the position of the line of the nodes, i.e. the intersection of the orbital plane with the mean ecliptic at the epoch (J2000), with respect to the reference $x$ axis pointing toward the Aries point $\Upsilon$.
Since, according to \rfr{opis}-\rfr{vobis}, the planet's coordinates $x,y,z$ enter the components of $\bds A_{n_{\rm X}}$, we also need the expressions for the unperturbed coordinates of the planet   \citep{Mont}
 \begin{eqnarray}
 x &=& r\left(\cos\Omega\cos u\ -\cos I\sin\Omega\sin u\right),\\ \nonumber\\
   y&=& r\left(\sin\Omega\cos u+\cos I\cos\Omega\sin u\right),\\ \nonumber\\
  z &=& r\sin I\sin u.
\end{eqnarray}
Thus, $A_r,A_t, A_n$ are non-linear functions of the three unknown parameters $\mathcal{K}_{\rm X}, \lambda_{\rm X},\beta_{\rm X}$ of X; by computing the averaged perihelion precessions induced by $\bds A_{\rm Hooke}$ and $\bds A_{n_{\rm X}}$ and comparing them with the estimated corrections $\Delta\dot\varpi$ to the usual perihelion precessions it is possible to have an upper bound on $\mathcal{K}_{\rm X}$, and, thus, a lower bound on $r_{\rm X}$ for given values of $M_{\rm X}$, as a function of  the position of X in the sky, i.e. $r_{\rm X}^{\rm min}=r_{\rm X}^{\rm min}(\lambda_{\rm X},\beta_{\rm X})$.
Relatively simple analytical expressions can be found only for particular positions of the body X; for example, by assuming $n_x=n_y=0$, i.e. by considering X located somewhere along the $z$ axis, it is possible to obtain
\eqi\left\langle\dot\omega\right\rangle_{n_{\rm X}}=\rp{9{\mathcal{K}_{\rm X}} \sqrt{1-e^2}\sin^2 I}{4n}\left(1 + \rp{5}{3}\cos 2\omega\right).\lb{proximat}\eqf In Table \ref{nemesis} we quote the lower bound on $r_{\rm X}=z_{\rm X}$ for several values of $M_{\rm X}$ obtained from  the maximum value of the estimated correction to the standard rate of the perihelion of Mars applied to \rfr{precis} and \rfr{proximat}.
In general, the problem must be tackled numerically; also in this case it turns out that the data from Mars yield the tightest constraints.
However, \rfr{precis} and \rfr{proximat} are already enough to show that  the effect of a body X could not be mimicked by a correction to the Sun's quadrupole mass moment $J_2$ accounting for its imperfect modelling in the EPM ephemerides. Indeed, it is easy to show  that the secular precession of the longitude of pericenter of a test body moving along an orbit with small eccentricity around an oblate body of equatorial radius $R$   is
\eqi \left\langle\dot\varpi\right\rangle_{J_2}=\rp{3}{2}n\left(\rp{R}{a}\right)^2\rp{J_2}{(1-e^2)^2}\left(\rp{5}{2}\cos^2 I -\cos I - \rp{1}{2}\right).\lb{quadru}\eqf
Now, let us consider the case in which X is directed along the $z$ axis and the planet P has $I=0$; only the retrograde Hooke-type precession of  \rfr{precis} would be non-vanishing, while \rfr{quadru} would induce a prograde rate.
\section{The three-dimensional spatial constraints on the location of X}\lb{megafig}

\begin{figure}
\includegraphics[width=\textwidth]{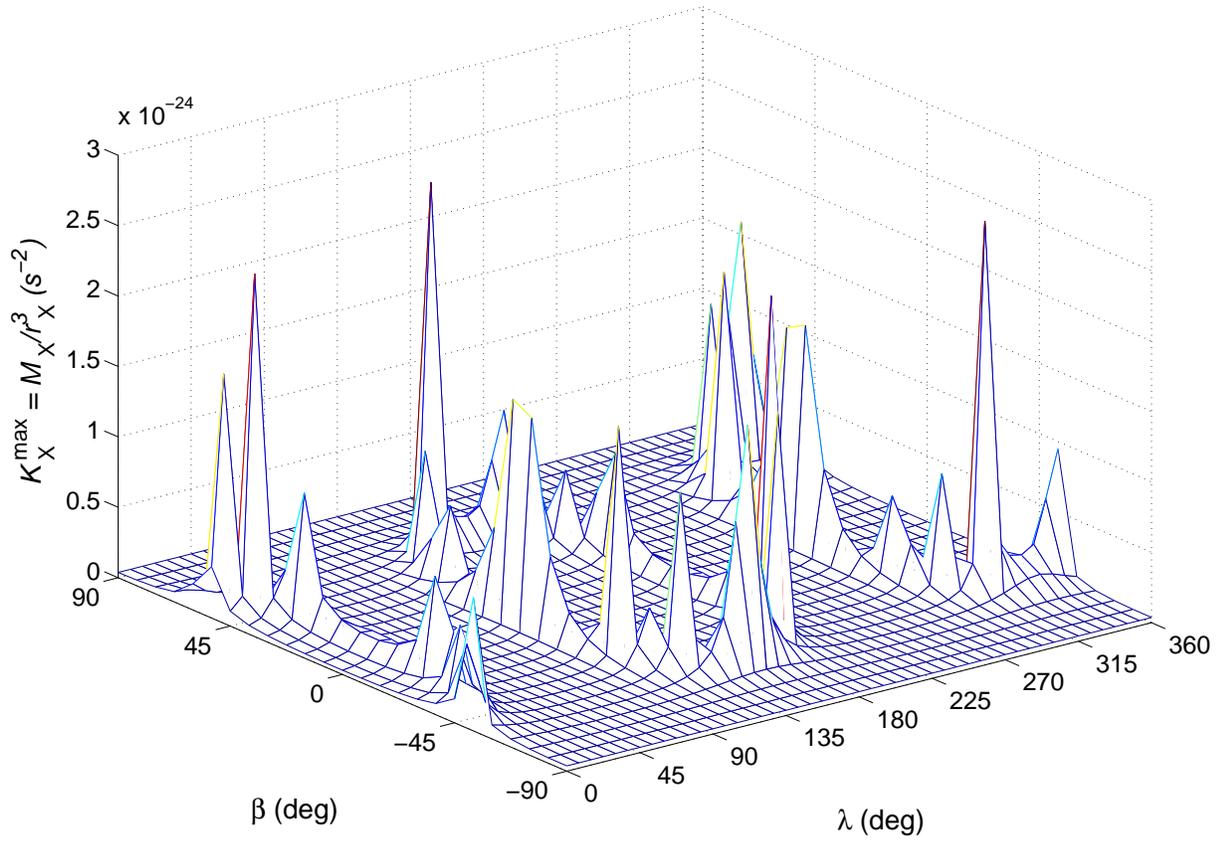}
\vspace*{8pt}
\caption{Maximum value, in s$^{-2}$, of  $\mathcal{K}_{\rm X}=GM_{\rm X}/r^3_{\rm X}$  as a function of the heliocentric longitude $\lambda$ and latitude $\beta$ of the putative body X; its upper bound is of the order of $3\times 10^{-24}$ s$^{-2}$. The perihelion of Mars has been used (Table \ref{chebolas}).}\label{NEMESIS_KAPPA}
\end{figure}

In Figure \ref{NEMESIS_KAPPA} we plot the maximum value of $\mathcal{K}_{\rm X}$, obtained from the perihelion of Mars which turns out to yield the most effective constraints, as a function of $\beta$ and $\lambda$.
It turns out that $\mathcal{K}_{\rm X}^{\rm max}$  is of the order of $10^{-24}$ s$^{-2}$.

The minimum distance $r_{\rm X}$ can be obtained for different values of $M_{\rm X}$ as a function of $\beta$ and $\lambda$ as well according to
\eqi r^{\rm min}_{\rm X}= \left(\rp{GM_{\rm X}}{\mathcal{K}^{\rm max}_{\rm X}}\right)^{1/3}.\eqf
We will consider the case of two rock-ice bodies with the masses of Mars and Earth, a giant planet with the mass of Jupiter, a brown dwarf with $M_{\rm X}=80 m_{\rm Jupiter}$, a red dwarf with $M_{\rm X}=0.5m_{\odot}$ and a Sun-mass body with $M_{\rm X}=M_{\odot}$, non necessarily an active main-sequence star.

It is also possible to visualize the forbidden region for X in the three-dimensional space by plotting its delimiting surface whose parameteric equations
are
\begin{eqnarray}
x &=& r^{\rm min}_{\rm X}(\beta,\lambda)\cos\beta\cos\lambda, \\ \nonumber \\
y &=& r^{\rm min}_{\rm X}(\beta,\lambda)\cos\beta\sin\lambda, \\ \nonumber \\
z &=& r^{\rm min}_{\rm X}(\beta,\lambda)\sin\beta.
\end{eqnarray}

\begin{figure}
\includegraphics[width=\textwidth]{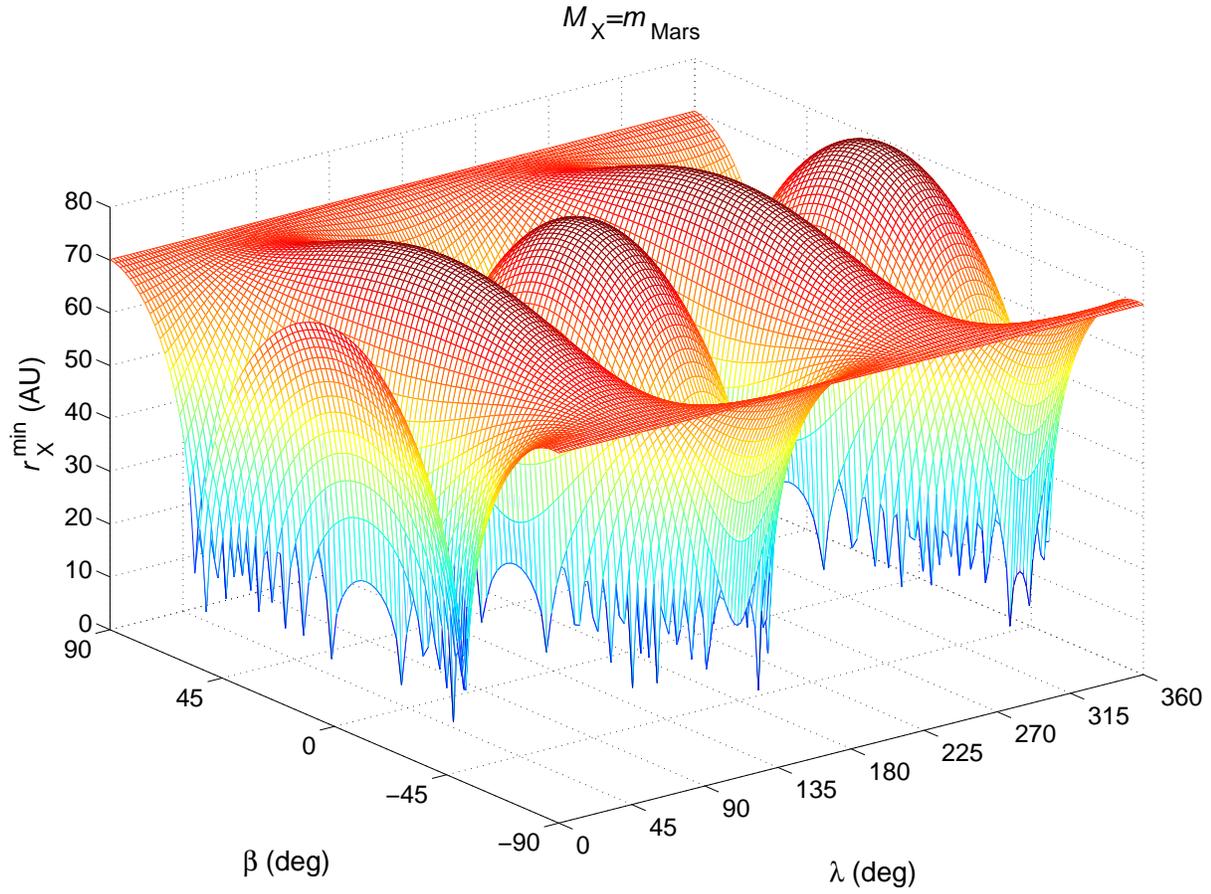}
\vspace*{8pt}
\caption{Minimum distance $r^{\rm min}_{\rm X}$ at which a  rock-ice planet of mass $M_{\rm X}= m_{\rm Mars}$ can exist  as a function of its heliocentric longitude $\lambda$ and latitude $\beta$. The perihelion of Mars has been used (Table \ref{chebolas}).}\label{NEMESISpolar_Marte}
\end{figure}

\begin{figure}
\includegraphics[width=\textwidth]{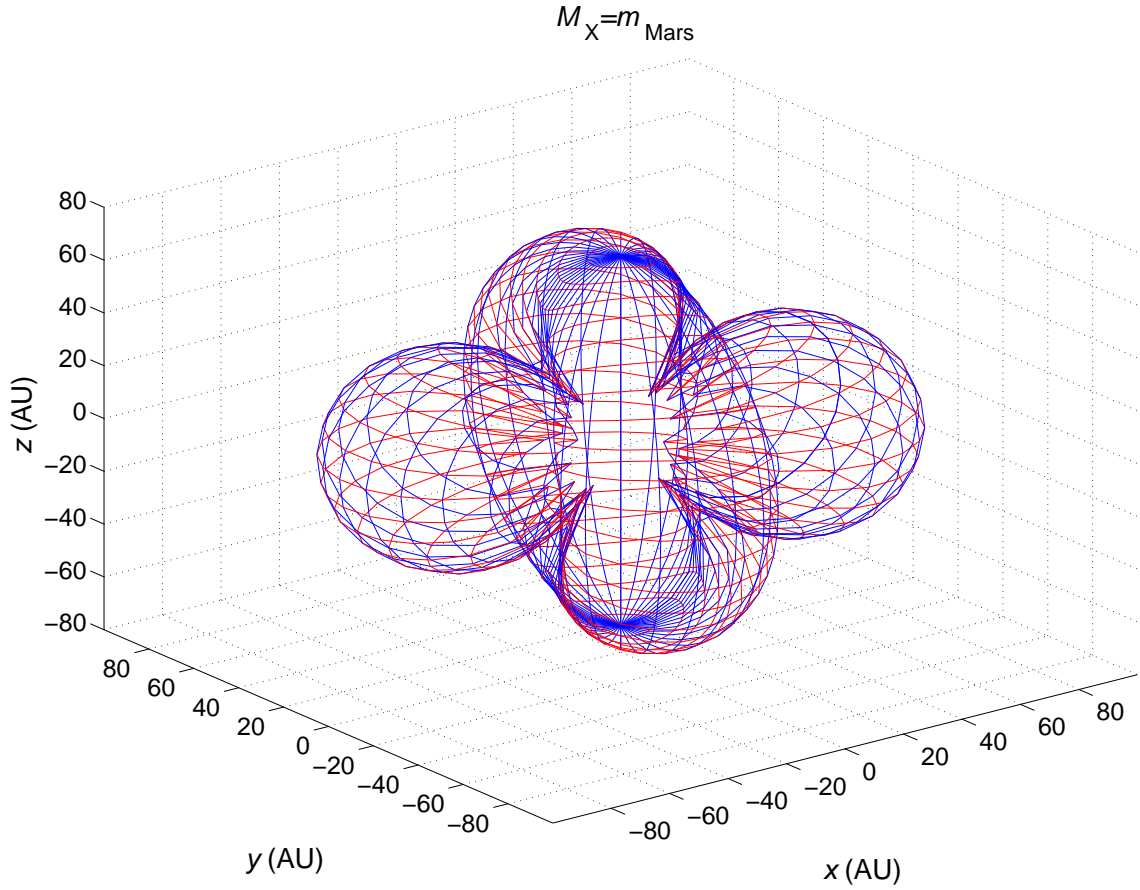}
\vspace*{8pt}
\caption{Surface delimiting the spatial region in which a rock-ice planet of mass $M_{\rm X}=m_{\rm Mars}$ can exist according to the dynamical constraints from the perihelion of Mars (Table \ref{chebolas}). The region inside the surface is forbidden: the region outside the surface is allowed. The red and blue lines correspond to constant values of $\beta$ and $\lambda$, respectively.}\label{NEMESIS3D_Marte}
\end{figure}

\begin{figure}
\includegraphics[width=\textwidth]{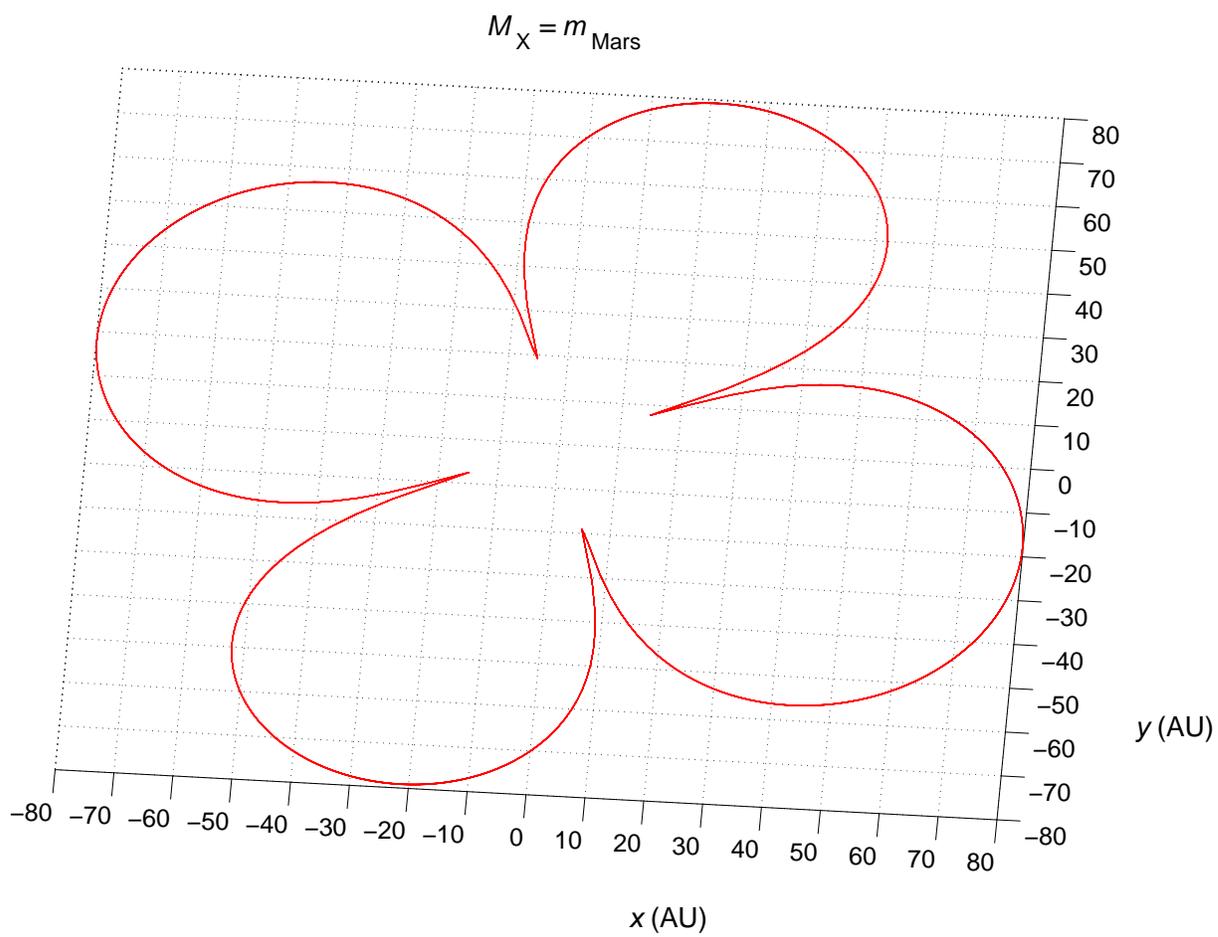}
\vspace*{8pt}
\caption{Ecliptic view: a Mars-sized rock-ice body can only exist outside the region delimited by the red contour. }\label{Mars_equatorial}
\end{figure}

In Figure \ref{NEMESISpolar_Marte}
we plot the minimum distance at which a body of mass $M_{\rm X}=m_{\rm Mars}$ can be found as a function of its latitude and longitude $\beta$ and $\lambda$.
The largest value is about 85 AU and occurs in the ecliptic plane ($\beta=0$) at about $\lambda = 0,60,150,250,345$ deg. Note that for $\beta=\pm 90$ deg, i.e. for X along the $z$ axis, the minimum distance is as in Table \ref{nemesis}, i.e.  70 AU.  For just a few positions in the sky such a Mars-sized body could be at no less than 20 AU.
In Figure \ref{NEMESIS3D_Marte} we depict the surface of minimum distance delimiting the spatial region in which such a body can exist according to the data from Mars. %
It has not a simple spherical shape, as it would have had if X exerted an isotropic force on Mars, and it has a precise spatial orientation with respect to the $\{x,y,z\}$ frame. The analytical results of Table \ref{nemesis} obtained with \rfr{proximat} in the case $\beta=\pm 90$ deg are confirmed by Figure \ref{NEMESISpolar_Marte} and Figure \ref{NEMESIS3D_Marte}. By slicing the surface with a vertical symmetry plane, it turns out that the shape of the central bun is rather oblate, approximately by $\approx 0.77$, contrary to the lateral lobes which are more spherical.  Moreover, the largest forbidden regions in the ecliptic plane are two approximately orthogonal strips with a length  of about 180 AU
(see Figure \ref{Mars_equatorial} depicting the situation in the ecliptic plane),
 while  in the vertical direction there is a strip approximately 120 AU long.

 Figures analogous to Figure \ref{NEMESISpolar_Marte}-Figure \ref{Mars_equatorial} hold for bodies with the mass of the Earth, Jupiter, a brown dwarf ($m=80m_{\rm Jupiter}$), a red dwarf with $M=0.5M_{\odot}$ and a Sun-sized body. The results of Table \ref{nemesis} concerning the position of X along the $z$ axis  are confirmed.  The same qualitative features of the case $M_{\rm X}=m_{\rm Mars}$ occur.  Concerning an Earth-sized body, it could mainly be found at no less than $147-175$ AU, with a minimum distance of 35 AU for just a few positions in the sky.  The minimum distance of a Jupiter-like mass is $1,006-1,200$ AU, with about $200$ AU in some points. For a brown dwarf ($M_{\rm X}=80\ m_{\rm Jupiter}$) the limiting distance is mainly $4,334-5,170$ AU, with a minimum of 861 AU at some positions, while for a red dwarf ($M_{\rm X}=0.5\ M_{\odot}$) it is $8,113-9,524$ AU, with a lowest value of 2,000 AU. Finally, a Sun-mass body cannot be located at less than $10,222-12,000$ AU for most of the sky positions, with $2,520$ AU at just a few points.
In the case of an Earth-sized body, the length  of the largest ecliptical forbidden strip turns out to be about 400 AU,
while the vertical one amounts to about 300 AU.

Concerning the strategy adopted so far, let us note that we  compared the estimated $\Delta\dot\varpi$ for the inner planets to their computed perihelion precessions due to X through \rfr{kazza}; in fact,  one should, in principle, also take into account the indirect effects of X on the inner planets of the solar system through its direct action on the outer planets. Indeed, if one of them is perturbed by X, its action on the inner planets will differ from the standard $N-$body one, fully modelled in the EPM ephemerides. To roughly evaluate such an effect, let us reason as follows.  The position of both X and of an outer planet  acted upon by X can approximately be considered  as constant in space with respect to a rocky planet over a typical orbital period $P_{\rm b}\lesssim 1$ yr. The maximum value of the disturbing acceleration imparted by X on a giant planet G like Jupiter will be of the order of
\eqi A_{\rm  GX}\leq 3\mathcal{K}_{\rm X}r_{\rm G}\approx 7\times 10^{-12}\ {\rm m\ s^{-2}};\eqf we used  $\mathcal{K_{\rm X}}=3\times 10^{-24}$ s$^{-2}$. $A_{\rm GX}$ will displace the position of G with respect to its usual position by an amount
\eqi \Delta r_{\rm G}\approx \rp{A_{\rm GX}}{2}P^2_{\rm b}\approx 3\times 10^3\ {\rm m}\eqf over $P_{\rm b}\lesssim 1$ yr during which $A_{\rm GX}$ can be considered as uniform.  As a consequence, the maximum extra-acceleration $\Delta A_{\rm PG}$ induced on an inner planet  P  by G during $P_{\rm b}$  can be evaluated as
\eqi\Delta A_{\rm PG}\approx 12\left(\rp{Gm_{\rm G}}{r_{\rm G}^4}\Delta r_{\rm G}\right)r\lesssim 2\times 10^{-15}\ {\rm m\ s^{-2}};\eqf we have just used \rfr{kazza}  adapted to the present case.
The perihelion precession induced by  $\Delta A_{\rm PG}$ can roughly be evaluated by dividing it by the orbital velocity of P, i.e. $an$, so that
\eqi\left\langle\dot\omega\right\rangle_{\rm G}\approx \rp{\Delta A_{\rm PG}}{an}\lesssim 10^{-5}\ {\rm arcsec\ cty^{-1}}.\eqf It is about one order of magnitude smaller than the present-day accuracy in knowing the extra-precessions of the perihelia of the inner planets, as shown by Table \ref{chebolas}.  Thus, we conclude that the indirect effects of X on the inner planets through its action on the giant ones can be neglected, as we did.

\section{Discussion and conclusions}\lb{fine}
We will, now, compare our dynamical constraints with other ones obtained with different methods.

 First, we compare the dynamical constraints of Table \ref{nemesis} and, more generally, of our full analysis with those obtainable as $r_{\rm X}=\sqrt{GM_{\rm X}/A_{\rm bary}}$ from the upper bound on the solar system barycenter's acceleration $A_{\rm bary}\leq 1.2\times 10^{-9}$ m s$^{-2}$ recently derived by    \citet{Zak} with an analysis of the timing data of several millisecond pulsars, pulsars in binary systems and pulsating white dwarf. Indeed, an acceleration exerted on the known SSB would affect the observed value of the rate of the period change of astronomical clocks such as pulsars and pulsating white dwarfs. Such a method is able to provide a uniform sensitivity over the entire sky. The pulsar-based, isotropic $1/r^2_{\rm X}$ constraints are summarized in Table \ref{nemesis2}.
With \virg{isotropic} we mean that the value of $A^{\rm max}_{\rm bary}$  used can be ruled out for $100\%$ of the sky at $95\%$ confidence level with practically all the methods used by \citet{Zak}; by using PSR B$1913+6$ the quoted acceleration is ruled out at $95\%$ confidence for $94\%$ of the sky.
They are not competitive with those of Table \ref{nemesis} and of the rest of our analysis for all the bodies considered.
However, according to \citet{noya}, future timing of millisecond pulsars looking for higher order pulsar period derivatives should be able to extend such limits to several thousand AU within a decade. A precise limit on the unmodeled relative acceleration between the solar system and PSR J0437-4715 has recently been obtained by \citet{Del} by  comparing  a VLBI-based measurement of the trigonometric parallax of PSR J0437-4715 to the kinematic distance obtained from pulsar timing, which is calculated from the pulsar's proper motion and apparent rate of change of orbital period. As a result, Jupiter-mass planets within 226 AU of the Sun in $50\%$ of the sky ($95\%$ confidence) are excluded.  Proposals to search for primordial black holes (PBHs) with the Square Kilometer Array by using modification of pulsar timing residuals when PBHs pass within about $1,000$ AU and impart impulse accelerations to the Earth  have been put forth by \citet{buchi}.

 Other acceleration-type $1/r^2_{\rm X}$  constraints were obtained by looking for a direct action of X on the Pioneer 10/11 spacecraft from an inspection of their tracking data \citep{And88}: the upper bound in the SSB acceleration obtained in this way is $4.2\times 10^{-10}$ m s$^{-2}$. This translate in a re-scaling of the values of Table \ref{nemesis2} by a factor $1.7$, still not competitive with ours.

\citet{Ho} looked for tidal-type  $1/r^3_{\rm X}$ constraints, like ours, with a dynamical approach based on  numerical simulations of the data of the four outer planets over the time span $1910-1990$. They found a relation among $M_{\rm X}$, in units of terrestrial masses, $r_{\rm X}$, in units of 0.1 kAU, and $\sigma_0$, which is the standard deviation of the assumed Gaussian random errors in $\lambda$ and $\beta$, in units of 0.1 arcsec; it is, for the ecliptic plane,
\eqi \left(\rp{r_{\rm X}}{100\ {\rm AU}}\right)\gtrsim \left[\left(\rp{M_{\rm X}}{m_{\oplus}}\right)\left(\rp{0.1\ {\rm arcsec}}{\sigma_0}\right)\rp{1}{(6-10)}\right]^{1/3}.\lb{Hog}\eqf
By assuming, rather optimistically, $\sigma_0\approx 0.01$ arcsec, \rfr{Hog} yields for $M_{\rm X}=m_{\oplus}$ $r_{\rm X}\geq 120$ AU, which is generally smaller than our limits for the ecliptic plane; a more conservative value of $\sigma_0\approx 0.1$ arcsec yields  $r_{\rm X}\geq 55$ AU.
It may be interesting to recall that \citet{coglio} predicted the existence of an alleged perturber of Uranus with $M_{\rm X}=0.6m_{\oplus}$ at a heliocentric distance of 44 AU.

Let us, now, focus on direct observational limits. A planetary body would reflect the solar light and, therefore, could be detected in the optical or near-infrared surveys. \citet{Tom} conducted an all-sky optical survey that, in the plane of the ecliptic, yielded the following constrain
\eqi \left(\rp{r_{\rm X}}{100\ {\rm AU}}\right)\gtrsim \left[\left(\rp{M_{\rm X}}{m_{\oplus}}\right)\rp{1}{3.4}\right]^{1/6},\eqf by assuming a visual albedo $p=0.02$ \citep{Ho}. For an Earth-sized body the limit is $r_{\rm X}\geq 81$ AU. Another optical survey yielding similar results was performed by \citet{Kow}.
In addition to the reflected sunlight,  also the X's own thermal emission would be detectable in the mid-to far-infrared. The $IRAS$ All-Sky Survey \citep{IRAS}  would have been able to discover a gas giant in the range 70 AU$\leq r_{\rm X} \leq 400$ AU, but without success, in agreement with our bounds for Jupiter-sized bodies.
A recent ecliptic survey was done by \citet{ECLI} with the Spacewatch telescope.  This survey was sensitive
to Mars-sized objects out to 300 AU and Jupiter-sized planets out to 1,200 AU; for low inclinations to the ecliptic,
 it ruled out more than one to two Pluto-sized objects out to 100 AU and one to two Mars-sized objects to
200 AU.
Concerning the case of a Sun-mass body, the presence of a main-sequence star above the hydrogen-burning limit was excluded within 1 pc by the all-sky synoptic Tycho-2 survey \citep{hog}.

Observational constraints on the properties of X might be obtained by sampling the astrometric position of background stars over the entire sky
with the future astrometric GAIA mission \citep{GAIA}. Indeed, the apparent motion  of X along the parallactic ellipse would deflect the angular position of distant stars due to the astrometric microlensing (\virg{induced parallax}).  A Jupiter-sized planet at 2,000 AU in the ecliptic plane could be detected by GAIA. Also the phenomenon of mesolensing \citep{MESO} could be used to gain information on possible planets at distances $> 1,000$ AU.   \citet{CMB} proposed to use the spectral distortion induced on the Cosmic Microwave Background (CMB) by putative distant masses to put constraints   on the physical properties and distances of them.  With the all-sky synoptic survey Pan-STARRS \citep{pan} massive planets such as Neptune would be undetectable through reflected sunlight beyond about 800 AU, while a body with $M_{\rm X}=0.1M_{\odot}$ would be undetectable for
$r_{\rm X}>2,000$ AU.

In conclusion, the dynamical constraints on a still undiscovered planet X in the outer regions of the solar system we dynamically obtained from the orbital motions of the inner planets of the solar system  with the extra-precessions of the perihelia estimated by Pitjeva with the EPM ephemerides
are tighter than those obtained from pulsar timing data analysis and the outer planets dynamics; moreover, they could be used in conjunction with the future planned all-sky surveys in the choice of the areas of the sky to be investigated. If and when other teams of astronomers will independently estimate their own correction to the standard perihelion precessions with different ephemerides it will be possible to repeat such tests. Moreover, a complementary approach to that presented here which could also be implemented consists of modifying the dynamical force models of the ephemerides by also including the action of X  on all the planets of the solar system, and repeating the global fitting procedure of the entire planetary data set estimating, among other parameters, also those which directly account for X itself and looking at a new set of planetary residuals.
%The dynamics of the inner planets of the solar system  allows to constrain the tidal parameter $|\mathcal{K}_{\rm X}|=GM_{\rm X}/r^3_{\rm X}$ of a %putative distant massive companion of the Sun (Planet X or Nemesis). The assumptions made are that its distance $r_{\rm X}$ is much larger than 1 AU %and that it can be considered constant over the typical orbital periods $P_{\rm b}\lesssim 1.5 $ yr of the rocky planets. It turns out that the %minimum distance at which a body with the same mass of the Earth could orbit is 130 AU, while for the mass of Mars it reduces to 62 AU; for the mass %of Jupiter it is 886 AU. By assuming it is a brown dwarf ($m\approx 75-80m_{\rm Jup}$), its minimum distance is about $3,800$ AU, while for a red %dwarf ($m\approx 0.075-0.5$ M$_{\odot}$) it is $3,793-7,139$ AU.  A star with the same mass of the Sun cannot be located at less than about $9,000$ %AU.
%Using the most recent determinations for the perihelion precession of Saturn allows to obtain slightly better constraints.
%Such constraints are of dynamical origin  and are in agreement with theoretical predictions involving the architecture of the Edgeworth-Kuiper Belt %independently obtained  with different approaches. Moreover, they yield tighter constraints than those obtainable from the upper bound on the %acceleration of the solar system  recently obtained from pulsar timing data.

\section*{Acknowledgments}
I gratefully thank C. Heinke and D. Ragozzine  for their important critical remarks.
%-----------------------------------------

\newpage

\begin{table}
\caption{Estimated corrections $\Delta\dot\varpi$, in milliarcsec cty$^{-1}$ (1 arcsec cty$^{-1} = 1.5\times 10^{-15}$ s$^{-1}$),  to the standard Newton/Einstein perihelion precessions
of the inner planets according to Table 3 of \citet{Pit05} (Mercury,
Earth, Mars). The result for Venus has been obtained by recently processing radiometric
data from Magellan spacecraft (E.V. Pitjeva, private communication, 2008).  The errors are not the formal, statistical ones. The SSB frame, assumed as inertial, i.e. without modelling the action of a putative body X, has been used.\label{chebolas}
}
\smallskip
\centering
\begin{tabular}{@{}cccc@{}}
\noalign{\smallskip}\hline\noalign{\smallskip}
 {Mercury} & {Venus} & {Earth} & {Mars}  \\
 \noalign{\smallskip}\hline\noalign{\smallskip}
 $-3.6\pm 5.0$ & $-0.4\pm 0.5$ & $-0.2\pm 0.4$ & $0.1\pm 0.5$\\
\noalign{\smallskip}\hline\noalign{\smallskip}
\end{tabular}
\end{table}
\begin{table}
\caption{Minimum heliocentric distance $r_{\rm X}$, in AU, at which a still unseen object having a mass $M_{\rm X}$ equal to that of the astronomical bodies listed here can be located along the $z$  axis ($\beta=\pm 90$ deg) according to \rfr{precis}, \rfr{proximat} and the maximum value of the extra-precession of the perihelion of Mars, according to  Table \ref{chebolas}. \label{nemesis}
}
\smallskip
\centering
\begin{tabular}{@{}cccccc@{}}
\noalign{\smallskip}\hline\noalign{\smallskip}
Mars & Earth & Jupiter & $M_{\rm X} = 80m_{\rm Jup}$ & $M_{\rm X} = \rp{M_{\odot}}{2}$ & Sun  \\
 \noalign{\smallskip}\hline\noalign{\smallskip}
$70$ & $147$ & $1,006$ & $4,336$ & $8,113$ & $10,222$\\
\noalign{\smallskip}\hline\noalign{\smallskip}
\end{tabular}
\end{table}

\begin{table}
\caption{Approximate minimum heliocentric distance $r_{\rm X}$, in AU, at which a still unseen object having a mass $M_{\rm X}$ equal to that of the astronomical bodies listed here can be located according to the limit $A_{\rm bary}=GM_{\rm X}/r_{\rm X}^2\leq 1.2\times 10^{-9}$ m s$^{-2}$ on the solar system barycenter's acceleration recently obtained  by \citet{Zak} from pulsar timing data. We used the maximum value $A_{\rm bary}/c=4\times 10^{-18}$ s$^{-1}$, where $c$ is the speed of light in vacuum, of the SSB acceleration allowed by pulsar timing data for about $100\%$ of the sky at $95\%$ level of confidence (Table 2 by \citet{Zak}).\label{nemesis2}
}
\smallskip
\centering
\begin{tabular}{@{}cccc@{}}
\noalign{\smallskip}\hline\noalign{\smallskip}
 Mars & Earth & Jupiter & Sun  \\
 \noalign{\smallskip}\hline\noalign{\smallskip}
 $1.4$ & $3.9$ & $68.6$ & $2,224$\\
\noalign{\smallskip}\hline\noalign{\smallskip}
\end{tabular}
\end{table}
\clearpage
\newpage
\clearpage
\newpage
\clearpage
\newpage
\clearpage
\newpage
\clearpage
\newpage

\end{document}